\documentclass[conference,letter]{IEEEtran}
\usepackage{comment}

\usepackage{geometry}
\usepackage[utf8]{inputenc} 
\usepackage[T1]{fontenc}
\usepackage{url}
\usepackage{ifthen}
\usepackage{cite}
\usepackage{amsmath}%
\usepackage{wasysym}%
\usepackage{amssymb}

\usepackage{mdwmath}
\usepackage{blindtext}
\usepackage{eqparbox}
\usepackage{lipsum}
\usepackage{xcolor}
\usepackage{algorithm}
\usepackage{enumerate,multirow}
\usepackage{algpseudocode}

\usepackage{times,amssymb,amsmath,amsfonts,nicefrac,color,bbm,mathrsfs,float}
\usepackage{multirow,tikz,graphicx}

\usepackage{xcolor}
\usepackage{comment}
\definecolor{MyGreen1}{RGB}{20,180,40}
\definecolor{MyBlue1}{RGB}{00,150,255}
\definecolor{MyGray1}{RGB}{200,200,200}

\usetikzlibrary{shapes.misc, positioning}
\usetikzlibrary{decorations.pathreplacing}

\usepackage{makecell}
\usetikzlibrary{fit}

\usetikzlibrary{arrows.meta,
	chains,
	decorations.pathreplacing,calligraphy,
	positioning,
	quotes,
	shapes.geometric,
	decorations.text
}

\tikzset{
BC/.style = {decorate,  
	decoration={calligraphic brace, amplitude=1.2mm,
		raise=1.2mm, mirror},
	thick, pen colour={black}
},
}

\tikzset{
	BC2/.style = {decorate,  
		decoration={calligraphic brace, amplitude=2mm,
			raise=1.5mm, mirror},
		thick, pen colour={black}
	},
}

\tikzset{
	block/.style    = {draw, thick, rectangle, text centered, align=center,  minimum width = 2em},
	sblock/.style      = {draw, thick, rectangle, minimum height = 2em,
		minimum width = 2em}, 
}

\tikzstyle{block} = [rectangle, draw,  text centered]
\tikzstyle{block} = [rounded corners]

\usepackage{siunitx}

\usepackage{calc}
\newcolumntype{M}[1]{>{\centering\arraybackslash}m{#1}}
\newcolumntype{N}{@{}m{0pt}@{}}

\usepackage{amsthm}

\DeclareMathAlphabet{\mathbfsl}{OT1}{ppl}{b}{it} 

\newcommand{\be}[1]{\begin{equation}\label{#1}}
\newcommand{\ee}{\end{equation}}

\newcommand{\Cref}[1]{Co\-ro\-lla\-ry\,\ref{#1}}

\IEEEoverridecommandlockouts
\begin{document}
\title{ProductAE: Toward Training Larger Channel Codes based on Neural Product Codes\vspace{-0.1cm}} 
\author{ \IEEEauthorblockN{Mohammad Vahid Jamali$^{\ast}$, Hamid Saber$^{\dagger}$, Homayoon Hatami$^{\dagger}$, and Jung Hyun Bae$^{\dagger}$}
\IEEEauthorblockA{$^{\ast}$Department of Electrical Engineering and Computer Science, University of Michigan, Ann Arbor, mvjamali@umich.edu\\
$^{\dagger}$Samsung Semiconductor, Inc., SOC Lab, \{hamid.saber, h.hatami, jung.b\}@samsung.com}
\vspace{-0.45cm}}	
\maketitle
\begin{abstract}
There have been significant research activities in recent years to automate the design of channel encoders and decoders via deep learning. Due the dimensionality challenge in channel coding, it is prohibitively complex to design and train relatively large neural channel codes via deep learning techniques. Consequently, most of the results in the literature are limited to relatively short codes having less than 100 information bits. In this paper, we construct ProductAEs, a computationally efficient family of deep-learning driven (encoder, decoder) pairs, that aim at enabling the training of relatively large channel codes (both encoders and decoders) with a manageable training complexity. We build upon the ideas from classical product codes, and propose constructing large neural codes using smaller code components. More specifically, instead of directly training the encoder and decoder for a large neural code of dimension $k$ and blocklength $n$, we provide a framework that requires training neural encoders and decoders for the code parameters $(n_1,k_1)$ and $(n_2,k_2)$ such that $n_1 n_2=n$ and $k_1 k_2=k$. Our training results show significant gains, over all ranges of signal-to-noise ratio (SNR), for a code of parameters $(225,100)$ and a moderate-length code of parameters $(441,196)$, over polar codes under successive cancellation (SC) decoder. Moreover, our results demonstrate meaningful gains over Turbo Autoencoder (TurboAE) and state-of-the-art classical codes. This is the first work to design product autoencoders and a pioneering work on training large channel codes.
\end{abstract}

\section{Introduction}\label{intro} 
Channel encoders and decoders are the essential components of any communication system that enable reliable communication by protecting the transmission of messages across a random noisy channel. 
Channel coding is one of the most principal problems in communication theory \cite{shannon1948mathematical,shannon1949communication}, and there are many landmark codes, such as Turbo codes \cite{berrou1993near}, low-density parity-check (LDPC) codes \cite{gallager1962low}, and polar codes \cite{arikan2009channel}, developed over decades of strong theoretical research. 

In recent years, there has been tremendous interest in the coding theory community to automate the design of channel encoders and decoders by incorporating deep learning methods \cite{kim2020physical,kim2018deepcode,jiang2019turbo,o2017introduction,makkuva2021ko,o2016learning,jamali2021Reed,ye2019circular,nachmani2016learning, gruber2017deep, nachmani2018deep, vasic2018learning, teng2019low, buchberger2020prunin,xu2017improved, cammerer2017scaling, bennatan2018deep, doan2018neural}. This is done by replacing the encoders and decoders (or some parts of the encoder and decoder architectures) with neural networks or some trainable models. 
The objectives and gains are vast, such as reducing the encoding and decoding complexity, improving the performance of classical channel codes, applications to realistic non-trivial channels and to emerging use cases, and designing universal decoders that simultaneously decode several codes, to mention a few. 

A major technical challenge here is the dimensionality issue that arises in channel coding context due to huge code spaces (there are $2^k$ distinct codewords for a binary linear code of dimension $k$). Therefore, it is prohibitively complex, if not impossible, to design and train relatively large neural channel encoders and decoders. In fact, only a small portion of all codewords will be seen during the training phase. Therefore, the trained models for the encoders and decoders may fail in generalizing to unseen codewords, which constitute the main portion of codewords for a relatively large $k$. Additionally, one needs to use huge networks with excessively large number of trainable parameters, which makes the training for larger $k$'s prohibitively complex.  Consequently, the achievements in the literature are mostly limited to relatively short codes having less than $100$ information bits.

Most of the literature focuses on decoding well-known codes using data-driven neural decoders \cite{nachmani2016learning, gruber2017deep, nachmani2018deep, vasic2018learning, teng2019low, buchberger2020prunin,xu2017improved, cammerer2017scaling, bennatan2018deep, doan2018neural}, and only a handful of works focus on discovering both encoders and decoders \cite{jiang2019turbo,o2016learning, o2017introduction,makkuva2021ko}.
A major challenge in jointly training the  encoder and decoder is to avoid local optima that may occur due to non-convex  loss functions.
It is worth mentioning that the pair of encoder and decoder together can be viewed as an over-complete autoencoder, 
where the goal is to find a higher dimensional representation of the input data  such that the original data can be reliably recovered from a noisy version of the higher dimensional representation  \cite{jiang2019turbo}.

  \begin{figure*}[t]
 	\centering
 \includegraphics[trim=2cm 12.4cm 2cm 12.7cm,width=6.8in]{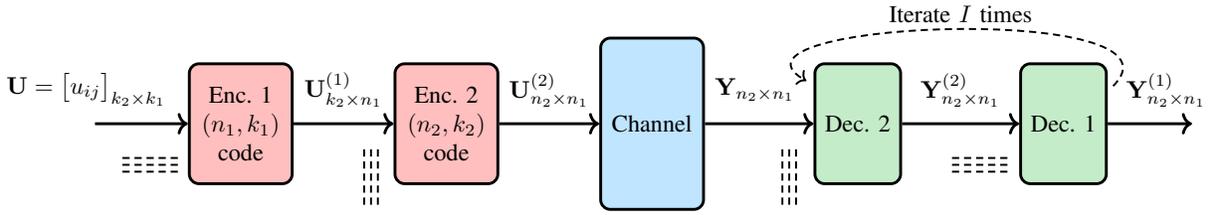}
 	\vspace{-0.25cm}
 	\caption{
 		Block diagram of a typical communication system incorporating two-dimensional (2D) product codes. Enc. 1 and Dec. 1 encode and decode rows of the 2D inputs while Enc. 2 and Dec. 2 encode and decode the columns of their input arrays.}
 	\label{fig1}
 	\vspace{-0.2cm}
 \end{figure*}

In this paper, we invent product autoencoders (ProductAEs) -- a new class of neural channel codes, i.e., a family of deep-learning driven (encoder, decoder) pairs -- with two major contributions: first, it is a pioneering work that targets enabling the training of large channel codes; and second, it is the first work on designing neural product codes.
To this end, we build upon the ideas from classical product codes, and propose constructing large neural codes using smaller code components. Particularly, instead of directly training the encoder and decoder for a large code of dimension $k$ and blocklength $n$, we provide a framework that requires training smaller neural encoder and decoder components of  parameters $(n_1,k_1)$, $(n_2,k_2)$,…,$(n_M,k_M)$ such that
$n_1n_2…n_M=n$ and $k_1k_2…k_M=k$ 
for some positive integer $M$.

Our training results for a ProductAE of parameters $(225,100)$, constructed based on the product of two smaller neural code components of parameters $(15,10)$, show significant performance gains compared to the polar code under successive cancellation (SC) decoding. More importantly, we demonstrate achieving the same gain for a moderate-length ProductAE of parameters $(441,196)$. This clearly establishes the generalization of our proposed architecture for training larger codes. 
Additionally, our results demonstrate meaningful gains over Turbo Autoencoder (TurboAE) and state-of-the-art classical codes. This important achievement is attained by applying innovative ideas from deep learning and intuitions from channel coding theory to further improve the training performance of our proposed neural product codes.

\section{Preliminaries}\label{prelim}
\subsection{Channel Coding}\label{prob_form}
Consider  transmission of a length-$k$ sequence of information bits $\mathbf{u}$ across a noisy channel. The channel encoder $\mathcal{E}(\cdot)$ maps $\mathbf{u}$ to a length-$n$ sequence of coded symbols $\mathbf{c}=\mathcal{E}(\mathbf{u})$ called a codeword. Here, $k$ and $n$ are the code dimension and blocklength, respectively, and the resulting code is denoted by an $(n,k)$ code. Also, the code rate is defined as $R=k/n$.
In the case of transmission over the additive white Gaussian noise (AWGN) channel, which is considered in this paper, the received noisy signal can be expressed as $\mathbf{y}=\mathbf{c}+\mathbf{n}$, where $\mathbf{n}$ is the channel noise vector whose components are Gaussian random variables with mean zero and variance $\sigma^2$. The ratio of the average energy per coded symbol to the noise variance is called the signal-to-noise ratio (SNR). In this paper,
we assume that the encoder satisfies a soft power constraint such that the average power per coded bit is $1$ (see Section \ref{subsec_norm}), and thus ${\rm SNR}=1/\sigma^2$.

On the other hand, the decoder $\mathcal{D}(\cdot)$ exploits the added redundancy in the encoder to estimate the transmitted message from the noisy received sequence as $\hat{\mathbf{u}}=\mathcal{D}(\mathbf{y})$. The main objective of channel coding is to increase the reliability by minimizing the error rate that is often measured by the 
bit error rate (BER) or block error rate (BLER), defined as ${\rm BER}=\frac{1}{k}\sum_1^k{\Pr}(\hat u_i \neq u_i)$ and ${\rm BLER}=\Pr(\hat{\mathbf{u}} \neq \mathbf{u})$, respectively.

\subsection{Product Codes}\label{PCs}
Product coding scheme is a powerful technique in constructing large channel codes, and building larger codes upon product codes can provide several advantages such as low encoding and decoding complexity, large minimum distances, and a highly parallelizable implementation \cite{elias1954error,pyndiah1998near,mukhtar2016turbo}. 
The encoding and decoding procedure of two-dimensional (2D) product codes is schematically shown in Fig. \ref{fig1}. Considering two codes $\mathcal{C}_1:(n_1,k_1)$ and $\mathcal{C}_2:(n_2,k_2)$, their product code that has the size $(n_1 n_2,k_1 k_2)$ is constructed in the following steps:
first, forming the length-$k_1 k_2$ information sequence as a $k_2\times k_1$ matrix $\mathbf{U}$; second,
encoding each row of $\mathbf{U}$ using $\mathcal{C}_1$ to get a matrix $\mathbf{U}^{(1)}$ of size $k_2\times n_1$ (or encoding each column using $\mathcal{C}_2$ to get a matrix of size $n_2\times k_1$); and third,
encoding each column of $\mathbf{U}^{(1)}$ using $\mathcal{C}_2$ to get a matrix $\mathbf{U}^{(2)}$ of size $n_2\times n_1$ (or encoding each row using $\mathcal{C}_1$ to get a matrix of size $n_2\times n_1$).
The decoding also proceeds the same way by first decoding each column and then each row (or vice versa) of the received noisy signal.

In general, to construct an $M$-dimensional product code $\mathcal{C}$, one needs to iterate $M$ codes $\mathcal{C}_1,\mathcal{C}_2,\cdots,\mathcal{C}_M$. In this case, each $m$-th encoder, $m=1,\cdots M$, encodes the $m$-th dimension of the $M$-dimensional input array.
Similarly, each $m$-th decoder decodes the noisy vectors on the $m$-th dimension of the received noisy signal.
 Assuming $\mathcal{C}_m:(n_m,k_m,d_m)$ with the generator matrix $\mathbf{G}_m$, where $d$ stands for the minimum distance, the parameters of the resulting product code $\mathcal{C}$ can be obtained as the product of the parameters of the component codes, i.e.,
\begin{align}
\mathcal{X}&=\prod_{m=1}^{M}\mathcal{X}_m,\hspace{1cm} \mathcal{X}\in\{n,k,d,R\},\label{X}\\
\mathbf{G}&=\mathbf{G}_1\otimes \mathbf{G}_2\otimes\cdots\otimes\mathbf{G}_M,\label{G}
\end{align}
where $\otimes$ is the Kronecker product operator.
A near-optimum iterative decoding algorithm for product codes built using linear block codes has been proposed in \cite{pyndiah1998near}, which is based on soft-input soft-output (SISO) decoding of the component codes. It has been shown in \cite{pyndiah1998near} and \cite{mukhtar2016turbo} that the decoding performance of product codes significantly improves using SISO decoders and also by applying several decoding iterations.
Interested readers are referred to \cite{mukhtar2016turbo} and the references therein for more details on product codes.  


\begin{figure*}[t]
	\centering
\includegraphics[trim=2cm 10cm 2cm 10.7cm,width=6.8in]{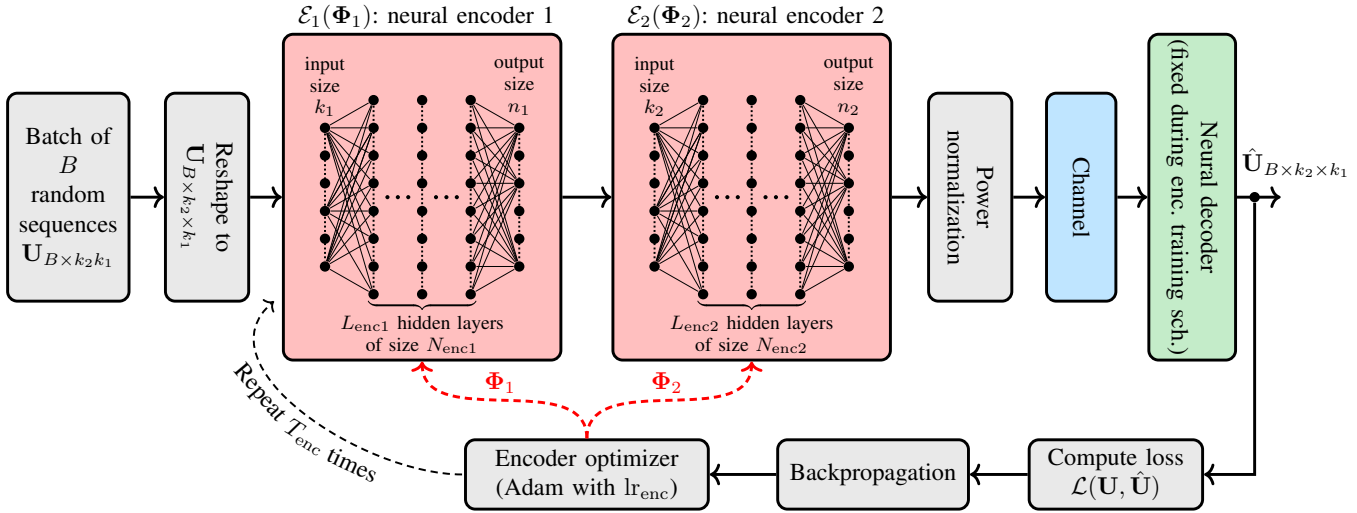}
	\vspace{-0.5cm}
	\caption{Encoder training schedule. The architecture of the neural product encoder is also depicted.}
	\label{fig2}	
	\vspace{-0.25cm}
\end{figure*}

\section{Product Autoencoder}\label{ProductAE}
\subsection{Proposed Architecture and Training}\label{proposed}
In this paper, we build upon 2D product codes. However, the proposed architectures and the methodologies naturally apply to higher dimensions. It is worth mentioning that the choice of 2D product codes balances a trade-off between the  complexity and performance. Indeed, from the classical coding theory perspective, although higher dimension product codes have lower encoding and decoding complexity for a given overall code size, they may result in inferior performance compared to lower dimension product codes due to exploiting the product structure to a greater extent (compared to a direct design).

In our ProductAE architecture, each encoder is replaced by a distinct neural network (NN). Also, assuming $I$ decoding iterations, each pair of decoders at each iteration is replaced by a distinct pair of NN decoders resulting in $2I$ NN decoders in total. Given that the ProductAE structure enables training encoders and decoders with relatively small code components, fully-connected NNs (FCNNs) are applied throughout the paper\footnote{We also explored convolutional NN (CNN)-based ProductAEs. However, we were able to achieve better training results with FCNN-based implementations.}.

ProductAE training is comprised of two main steps: $(i)$ decoder training schedule; and $(ii)$ encoder training schedule. More specifically, during each training epoch, we first train the decoder several times while keeping the encoder network fixed, and then we train the encoder multiple times while keeping the decoder weights unchanged. In the following, we first present the encoder architecture and describe its training schedule, and then we do the same for the decoder.

\subsubsection{Encoder Architecture and Its Training Schedule}\label{sec_enc}
As shown in Fig. \ref{fig2}, we replace the two encoders by two FCNNs, namely $\mathcal{E}_1(\mathbf{\Phi}_1)$ and $\mathcal{E}_2(\mathbf{\Phi}_2)$, each parametrized with a set of weights $\mathbf{\Phi}_1$ and $\mathbf{\Phi}_2$, respectively. Each $j$-th encoder, $j=1,2$, has a input size of $k_j$, output size of $n_j$, and $L_{{\rm enc}j}$ hidden layers of size $N_{{\rm enc}j}$. Upon receiving a batch of $k_2\times k_1$ arrays of information bits, the first encoder maps each length-$k_1$ row to a length-$n_1$ real-valued vector, resulting in a tensor $\mathbf{U}^{(1)}_{B\times k_2 \times n_1}$. The second NN encoder then maps each real-valued length-$k_2$ vector in the second dimension of $\mathbf{U}^{(1)}$ to a length-$n_2$ real-valued vector. Note that the mappings here are, in general, nonlinear mappings, and the resulting code is a nonlinear and non-binary code. At the end, the power normalization, to be clarified in Section \ref{impl_detl}, will be applied to the codewords.

The encoder training schedule at each epoch is as follows. First, a batch of $B$ length-$k_1k_2$ binary information sequences will be reshaped to a tensor $\mathbf{U}_{B\times k_2 \times k_1}$ to fit the NN encoders. After encoding the input, the real-valued codewords will be passed through the AWGN channel and then will be decoded using the NN decoder (explained in Section \ref{sec_dec}), that is fixed throughout the encoder training schedule, to get the batch of decoded codewords $\hat{\mathbf{U}}_{B\times k_2k_1}$ (after appropriate reshaping). By computing the loss between the transmitted and decoded sequences $\mathcal{L}(\mathbf{U},\hat{\mathbf{U}})$ and backpropagating the loss to compute its gradients, the encoder optimization takes a step to update the weights of the NN encoders $\mathcal{E}_1(\mathbf{\Phi}_1)$ and $\mathcal{E}_2(\mathbf{\Phi}_2)$. This procedure will be repeated $T_{\rm enc}$ times while each time only updating the encoder weights for a fixed decoder model.

\begin{figure*}[t]
	\centering
	\includegraphics[trim=2cm 10.5cm 2cm 10.8cm,width=6.8in]{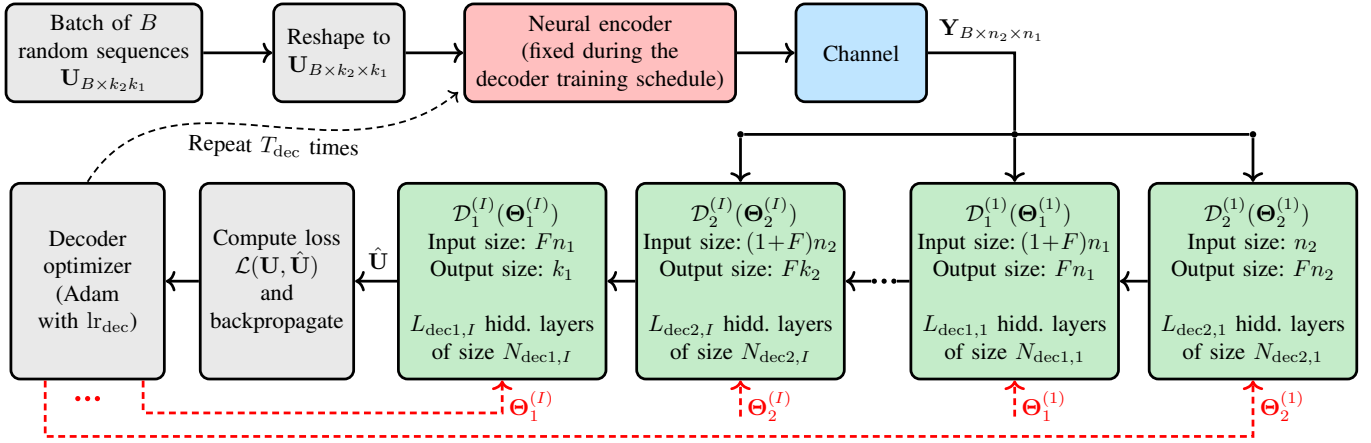}
	\vspace{-0.6cm}
	\caption{Decoder training schedule. The architecture of the neural product decoder is also depicted. $I$ iterations of the decoder are replaced by $2I$ neural decoders $\mathcal{D}^{(i)}_j(\mathbf{\Theta}^{(i)}_j)$, $i=1,...,I$, $j=1,2$, parametrized by the set of trainable weights $\mathbf{\Theta}^{(i)}_j$.}
	\label{fig3}
	\vspace{-0.25cm}
\end{figure*}

\subsubsection{Decoder Architecture and Its Training Schedule}\label{sec_dec}

As shown in Fig. \ref{fig3}, we replace the pair of decoders at each $i$-th iteration, $i=1,\cdots I$, by a pair of distinct FCNNs resulting in $2I$ NN decoders $\mathcal{D}^{(i)}_j$, $j=1,2$, each parametrized by a set of weights $\mathbf{\Theta}^{(i)}_j$. The architecture of each FCNN decoder is depicted in Fig. \ref{fig3}. Each NN decoder $\mathcal{D}^{(i)}_j$ has $L_{{\rm dec}j,i}$ hidden layers of size $N_{{\rm dec}j,i}$. Generally speaking, the first $I-1$ pairs of NN decoders work on input and output sizes of the length of coded bits $n_j$, while the last pair reverts the encoding operation by reducing the lengths from $n_j$ to $k_j$. Also, as further clarified in Section \ref{modif}, some of the decoders may take multiple length-$n_j$ vectors as input, and output multiple copies of length-$n_j$ vectors. 

During the decoder training schedule, at each epoch, first a batch of $B$ reshaped binary information sequences will be encoded using the fixed NN encoder. The real-valued codewords will be passed through the AWGN channel, with a carefully selected (range of) decoder training SNR, and then will be decoded using the NN decoder to get the batch of decoded codewords $\hat{\mathbf{U}}_{B\times k_2k_1}$. After computing the loss and its gradients, the decoder optimizer then updates the weights of the NN decoders. This procedure is repeated $T_{\rm dec}$ times while each time only updating the decoder weights for a fixed encoder model.

\subsection{Additional Modifications}\label{modif}
In this section, we briefly present additional modifications applied to the NN decoder architecture.

\subsubsection{Channel Output Injection}\label{chnl_inj}
Following the structure of the SISO decoder for classical product codes \cite{pyndiah1998near}, in addition to the output from the previous decoder, we also pass the channel output to all NN decoders. In this case, the first decoder $\mathcal{D}_2^{(1)}$ only takes the channel output as the input.
All the next $2I-2$ decoders take the channel output besides the output of the previous decoder, while the last decoder $\mathcal{D}_1^{(I)}$ only takes the output of the previous decoder as input (due to some issues with the size of the tensors being concatenated at the input of this decoder).
Our initial results show that this method \textit{alone} is not able to make a considerable improvement in the training.
However, the combination of this method with the other modifications presented in this section is promising.

\subsubsection{Adding Feature Size}\label{feature_size}
As shown in Fig. \ref{fig3}, each NN decoder, except the last one, outputs $F$ vectors as the soft information, where $F$ denotes the feature size. This is equivalent to saying that each NN decoder outputs $F$ different  estimates of the soft information instead of one. This is done by increasing the output size of the FCNNs. These estimates will then be carefully reshaped and processed before feeding the next decoder such that $F$ properly formed vectors are given to the next decoder (increasing the input size of FCNNs).

\subsubsection{Subtraction of Soft Information}\label{subtrct}
We input the difference of the soft input from the soft output of a decoder to the next decoder, i.e., the increments of the soft information is given to the next decoder. To this end,
the first decoder $\mathcal{D}_2^{(1)}$ only takes the channel output. Also, the second decoder directly receives the output of the first decoder since there was no soft input to $\mathcal{D}_2^{(1)}$. All the next decoders, except the last one (due to dimensionality issue), take the increments of soft information.

We carried out extensive experiments to investigate the impacts of the aforementioned modifications. Our training results demonstrate that the combination of all these three methods together significantly improves the training performance.

\subsubsection{Other Modifications}\label{othr_mods} Besides the modifications detailed above, we also explored some other modifications to our ProductAE architecture, some of which helped in improving the training performance and some did not.
For example, we explored adding an $l_2$ regularizer to the loss function of the encoder. We also investigated replacing the subtraction operation between the soft input and output of NN decoders (see Section \ref{subtrct}) with some FCNNs. 
To do so, we defined distinct FCNNs of input size $2$ and output size $1$, each doing an element-wise operation between an element of the soft input and an element of the soft output to substitute their difference. Through extensive experiments, however, we were not able to improve upon our best training results with these two modifications. One particularly useful method though was loading the best model of an experiment after the saturation of the training loss versus the number of epochs, and then running a few number of epochs with a significantly large batch size. In Section \ref{impl_detl}, we discuss how we handle training with very large batch sizes while avoiding memory shortage.

\subsection{Further Implementation Details}\label{impl_detl}
In this section, we briefly discuss some other aspects of the implementations that have not been covered above.

\subsubsection{Training with Very Large Batch Sizes}\label{subsec_B}
It is well known that increasing the batch size can significantly improve the training performance by averaging out the channel noise \cite{jiang2019turbo}. Throughout our experiments, we used relatively large batches of size at least $B=5,000$. Also, as discussed in Section \ref{impl_detl}, several epochs with much larger batch sizes are applied after the saturation of the model. In order to accommodate large enough batch sizes, beyond the memory shortage, we run $l$ smaller batches of size $B_s$ while accumulating the gradients of loss without updating the optimizer across these batches. The overall operation is then equivalent to having a single large batch of size $lB_s$.

\subsubsection{Optimizers, Activation Functions, and Loss Function}\label{subsec_opt}
\begin{itemize}
	\item ``Adam'' optimizer is chosen with learning rate ${\rm lr_{enc}}$ for the encoder and ${\rm lr_{dec}}$ for the decoder.
	\item ``SELU'' (scaled exponential linear units) is applied as the activation function to all hidden layers. No activation function is applied to the output layer of any of the NN encoders and decoders.
	\item ``BCEWithLogitsLoss()'' is adopted as the loss function, which combines a ``sigmoid layer'' and the binary cross-entropy (BCE) loss in one single class in a more numerically stable fashion.
\end{itemize}

\subsubsection{Encoder and Decoder Training Schedules and SNRs}\label{subsec_snr} The encoder and decoder training schedules were discussed in Section \ref{proposed}. Throughout our experiments, we used $T_{\rm enc}=100$ and $T_{\rm dec}=500$. Also, at each encoder or decoder training iteration, the training SNR (equivalently, the noise variance of the channel) is carefully chosen. In particular, we used a single-point of $\gamma$ \si{dB} for the encoder training SNR but a range of $[\gamma-2.5,\gamma+1]$ \si{dB} for the decoder training SNR. Therefore,
during each encoder training iteration, a single SNR is used to generate all noise vectors of the batch. However, during each iteration of the decoder training schedule, $B$ random values are picked uniformly from the interval $[\gamma-2.5,\gamma+1]$ \si{dB} to generate $B$ noise vectors of appropriate variances.

\subsubsection{Encoder Power Normalization}\label{subsec_norm}
In order to ensure that the average power per coded symbol is equal to one and thus the average SNR is $1/\sigma^2$, the length-$n$ real-valued encoded sequences $\mathbf{c}=(c_1,c_2,…,c_n)$ are normalized as follows.
\begin{align}\label{norm}
\mathbf{c}'=\sqrt{n}\mathbf{c}/||\mathbf{c}||_2.
\end{align}

\section{Experiments}\label{experiments}
In this section, we first present the performance results for two ProductAEs of rate $4/9$, namely, ProductAE $(15,10)^2$ and a moderate-length ProductAE $(21,14)^2$, where $(n,k)^2$ denotes the product of two $(n,k)$ codes. We then compare the ProductAE performance with TurboAE and also state-of-the-art classical channel codes.
Our trained models are obtained with $B=5000$, ${\rm lr_{enc}}={\rm lr_{dec}}=2\times 10^{-4}$, $I=4$, $F=3$, and $\gamma=3$ dB. All FCNNs have $7$ hidden layers except the last pair of decoders which have $9$ layers. Also, the size of all hidden layers is set to $N_{\rm enc}=200$ for the encoder FCNNs and $N_{\rm dec}=250$ for the decoder FCNNs.

\subsection{ProductAE $(15,10)^2$}\label{10_15}
\begin{figure}[t]
	\centering
	\includegraphics[trim=0.3cm 0.5cm 0 0,width=3.6in]{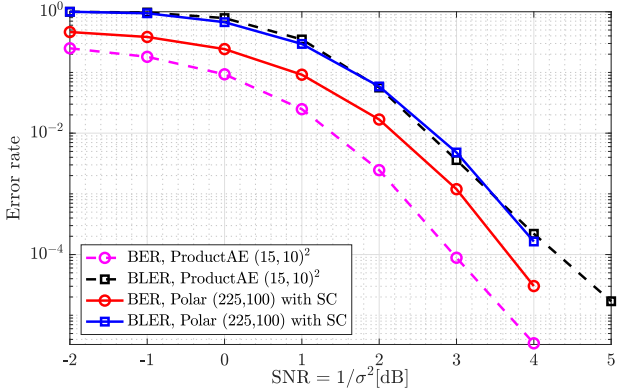}
	\caption{Testing result of ProductAE $(15,10)^2$ compared to polar code $(225,100)$ under SC decoding.}
	\label{fig4}
	\vspace{-0.2cm}
\end{figure}

Fig. \ref{fig4} presents the performance of ProductAE $(15,10)^2$, where a single model is tested across all ranges of SNRs. As a benchmark, the ProductAE performance is compared to a polar code of the same length, i.e., with parameters $(225,100)$, under SC decoding. This polar code is obtained by puncturing $31$ coded bits of a $(256,100)$ polar code at random. As seen, our trained ProductAE $(15,10)^2$ beats the BER performance of the equivalent polar code with a significant margin over all ranges of SNR. Also, even though the loss function is a surrogate of the BER, our ProductAE is able to achieve almost the same BLER as the polar code.
\subsection{ProductAE $(21,14)^2$}\label{14_21}
\begin{figure}[t]
	\centering
	\includegraphics[trim=0.3cm 0.5cm 0 0,width=3.6in]{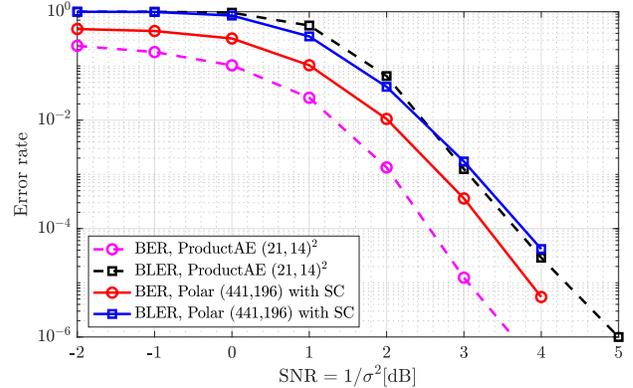}
	\caption{Testing result of ProductAE $(21,14)^2$ compared to polar code $(441,196)$ under SC decoding.}
	\label{fig5}
	\vspace{-0.2cm}
\end{figure}

Fig. \ref{fig5} compares the performance of the trained ProductAE $(21,14)^2$ with a polar code of parameters $(441,196)$, that is obtained by puncturing $71$ coded bits of a $(512,196)$ polar code at random, under SC decoding. It is truly remarkable that the moderate-length ProductAE $(21,14)^2$ beats the BER performance of the equivalent polar code with a large margin over all ranges of SNR while also maintaining almost the same BLER performance. To the best of our knowledge, this is the first-ever work to report successful training of pure neural (encoder, decoder) pairs, i.e., a completely new class of nonlinear neural codes paired with efficient decoders, over such relatively large code dimensions.

\subsection{Comparisons with TurboAE}\label{turboAE}
In this section, we compare the performance of ProductAEs with rate-1/3 TurboAEs $(300,100)$ and $(600,200)$. TurboAE $(300,100)$ is directly trained \cite{jiang2019turbo} while TurboAE $(600,200)$ result is obtained by testing a model trained for TurboAE $(300,100)$. Given that the trained ProductAEs have a different rate of $4/9$ compared to TurboAEs, we compare their performance in terms of the energy-per-bit to the noise ratio, denoted by $E_b/N_0$ in this paper, to be discussed in the following.

\begin{figure}[t]
	\centering
	\includegraphics[trim=0.3cm 0.5cm 0 0,width=3.6in]{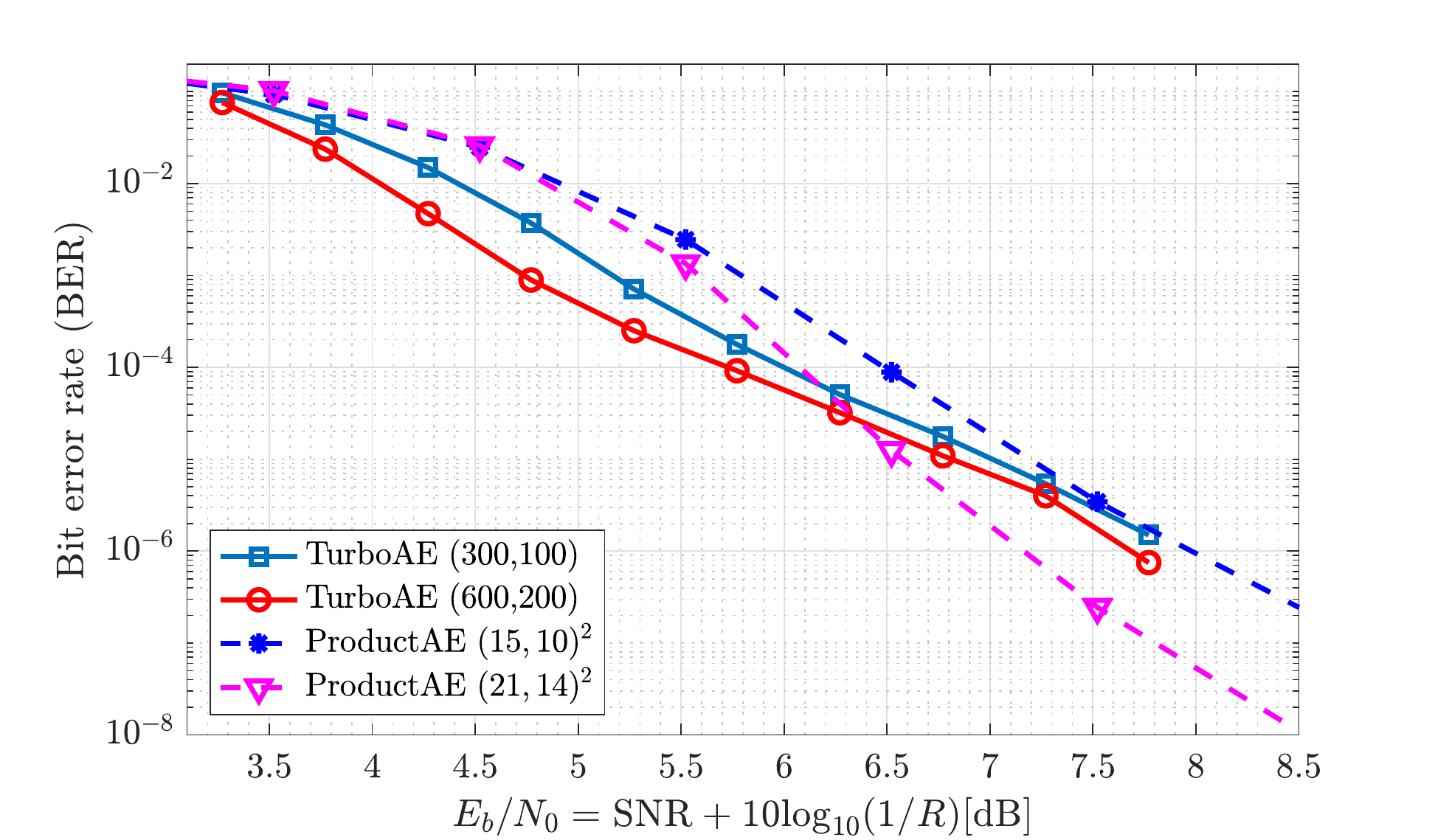}
	\caption{Performance comparison of ProductAE and TurboAE.}
	\label{fig6}
	\vspace{-0.2cm}
\end{figure}

One way to compare the performance of two codes of different rates is by looking at the $E_b/N_0$ instead of SNR. Note that $E_b/N_0={\rm SNR}/R$,\footnote{Note that our definitions of ${\rm SNR}$ and $E_b/N_0$ as $1/\sigma^2$ and $1/R\sigma^2$, respectively, are slightly different from the definitions $1/2\sigma^2$ and $1/2R\sigma^2$, respectively, used in some other works \cite{jamali2021Reed}.} i.e., $E_b/N_0={\rm SNR}+10\log_{10}(1/R)$ in \si{dB}. Fig. \ref{fig6} compares the performance of ProductAEs with TurboAEs in $E_b/N_0$. It is observed that the ProductAE $(21,14)^2$ beats the performance of TurboAE $(600,200)$ with a good margin at relatively low BERs. For example, there are $0.5$ \si{dB} and $0.2$ \si{dB} gains at the BERs of $10^{-6}$ and $10^{-5}$, respectively. Also,  ProductAE $(15,10)^2$ approaches the performance of TurboAE $(300,100)$ at high $E_b/N_0$'s. 

Note that this comparison is in favor of TurboAE as it has a larger block-length than the ProductAE counterparts, and it is well known that the performance improves for larger length codes. Therefore, a fair comparison requires looking at TurboAEs of the same blocklength as our ProductAEs, e.g., by puncturing $75$ and $159$ coded bits of TurboAEs $(300,100)$ and $(600,200)$, respectively, which degrades the TurboAEs performance compared to the results in Fig. \ref{fig6}.

\subsection{Comparisons with Classical Codes}\label{classical_codes}
\begin{figure}[t]
	\centering
	\includegraphics[trim=0.3cm 0.5cm 0 0,width=3.6in]{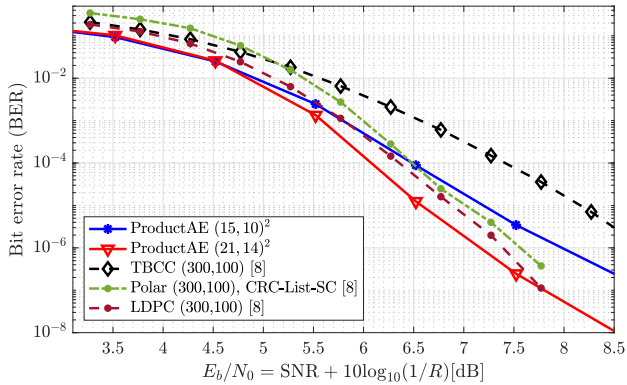}
	\caption{Performance comparison of ProductAE with classical codes.}
	\label{fig7}
	\vspace{-0.2cm}
\end{figure}

Fig. \ref{fig7} compares the performance of our ProductAEs with state-of-the-art classical codes, namely polar under CRC-List-SC decoder, LDPC, and tail-biting convolutional code (TBCC). All these three codes have parameters $(300,100)$, and their performance are directly extracted from \cite[Fig. 1]{jiang2019turbo}. Remarkably,  ProductAE $(15,10)^2$ beats the performance of TBCC with a good margin over all ranges of $E_b/N_0$ and that of polar and LDPC codes for BER values of larger than $10^{-4}$. Similar to TurboAE, the comparison here is against ProductAE, since a fair comparison requires reducing the blocklength of the considered classical codes to $225$ bits, e.g., through puncturing $75$ bits. Note that ProductAE $(21,14)^2$, with less than $1.5$ times larger blocklength but almost twice larger code dimension, is able to outperform the state-of-the-art classical codes, with a good margin, over all BER ranges of interest.

\section{Conclusions}\label{concl}
In this paper, we presented training ProductAEs motivated by the goal of training larger channel codes based on smaller code components. This work is the first to explore training product codes, and is a pioneering work in training large channel codes. Although we demonstrated excellent performance gains, compared to polar code and TurboAE, as well as state-of-the-art classical codes, via training ProductAEs of dimensions 100 and 196, the architecture and training procedure are general and can be applied to train larger ProductAEs. Given that the ProductAE enables a low encoding and decoding complexity, flexible code rate and length, parallel implementation, etc., the achieved performance gains are truly remarkable, and call for extensive future research. This work paves the way toward designing larger neural codes that beat the state-of-the-art performance with potential applications in the design of the next generations of wireless networks.



\begin{thebibliography}{10}
	\providecommand{\url}[1]{#1}
	\csname url@samestyle\endcsname
	\providecommand{\newblock}{\relax}
	\providecommand{\bibinfo}[2]{#2}
	\providecommand{\BIBentrySTDinterwordspacing}{\spaceskip=0pt\relax}
	\providecommand{\BIBentryALTinterwordstretchfactor}{4}
	\providecommand{\BIBentryALTinterwordspacing}{\spaceskip=\fontdimen2\font plus
		\BIBentryALTinterwordstretchfactor\fontdimen3\font minus
		\fontdimen4\font\relax}
	\providecommand{\BIBforeignlanguage}[2]{{%
			\expandafter\ifx\csname l@#1\endcsname\relax
			\typeout{** WARNING: IEEEtran.bst: No hyphenation pattern has been}%
			\typeout{** loaded for the language `#1'. Using the pattern for}%
			\typeout{** the default language instead.}%
			\else
			\language=\csname l@#1\endcsname
			\fi
			#2}}
	\providecommand{\BIBdecl}{\relax}
	\BIBdecl
	
	\bibitem{shannon1948mathematical}
	C.~E. Shannon, ``A mathematical theory of communication,'' \emph{The Bell
		system technical journal}, vol.~27, no.~3, pp. 379--423, 1948.
	
	\bibitem{shannon1949communication}
	{C. E. Shannon}, ``Communication in the presence of noise,'' \emph{Proceedings
		of the IRE}, vol.~37, no.~1, pp. 10--21, 1949.
	
	\bibitem{berrou1993near}
	C.~Berrou, A.~Glavieux, and P.~Thitimajshima, ``Near {Shannon} limit
	error-correcting coding and decoding: {Turbo}-codes. 1,'' in \emph{Proc. IEEE
		Int. Conf. Commun. (ICC)}, vol.~2, May 1993, pp. 1064--1070.
	
	\bibitem{gallager1962low}
	R.~Gallager, ``Low-density parity-check codes,'' \emph{IRE Trans. Inf. Theory},
	vol.~8, no.~1, pp. 21--28, 1962.
	
	\bibitem{arikan2009channel}
	E.~Arikan, ``Channel polarization: A method for constructing capacity-achieving
	codes for symmetric binary-input memoryless channels,'' \emph{IEEE Trans.
		Inf. Theory}, vol.~55, no.~7, pp. 3051--3073, 2009.
	
	\bibitem{kim2020physical}
	H.~Kim, S.~Oh, and P.~Viswanath, ``Physical layer communication via deep
	learning,'' \emph{IEEE J. Sel. Areas Inf. Theory}, vol.~1, no.~1, pp.
	194--206, 2020.
	
	\bibitem{kim2018deepcode}
	H.~Kim, Y.~Jiang, S.~Kannan, S.~Oh, and P.~Viswanath, ``Deepcode: feedback
	codes via deep learning,'' \emph{IEEE J. Sel. Areas Inf. Theory}, vol.~1,
	no.~1, pp. 194--206, 2020.
	
	\bibitem{jiang2019turbo}
	Y.~Jiang, H.~Kim, H.~Asnani, S.~Kannan, S.~Oh, and P.~Viswanath, ``Turbo
	autoencoder: Deep learning based channel codes for point-to-point
	communication channels,'' \emph{Proc. Adv. Neural Inf. Process. Syst.
		(NeurIPS)}, vol.~32, pp. 2758--2768, 2019.
	
	\bibitem{o2017introduction}
	T.~O’shea and J.~Hoydis, ``An introduction to deep learning for the physical
	layer,'' \emph{$\!$IEEE $\!$Trans. $\!$Cognit. $\!$Commun. $\!$Netw.}, vol.
	$\!$3, no. $\!$4, pp. $\!$563--575, $\!$2017.
	
	\bibitem{makkuva2021ko}
	A.~V. Makkuva, X.~Liu, M.~V. Jamali, H.~Mahdavifar, S.~Oh, and P.~Viswanath,
	``{KO} codes: inventing nonlinear encoding and decoding for reliable wireless
	communication via deep-learning,'' in \emph{Proc. Int. Conf. Mach. Learn.
		(ICML)}.\hskip 1em plus 0.5em minus 0.4em\relax PMLR, 2021, pp. 7368--7378.
	
	\bibitem{o2016learning}
	T.~J. O'Shea, K.~Karra, and T.~C. Clancy, ``Learning to communicate: Channel
	auto-encoders, domain specific regularizers, and attention,'' in \emph{Proc.
		IEEE Int. Symp. Signal Process. Inf. Technol.}, 2016, pp. 223--228.
	
	\bibitem{jamali2021Reed}
	M.~V. Jamali, X.~Liu, A.~V. Makkuva, H.~Mahdavifar, S.~Oh, and P.~Viswanath,
	``{Reed-Muller} subcodes: Machine learning-aided design of efficient soft
	recursive decoding,'' in \emph{Proc. IEEE Int. Symp. Inf. Theory (ISIT)},
	2021, pp. 1088--1093.
	
	\bibitem{ye2019circular}
	H.~Ye, L.~Liang, and G.~Y. Li, ``Circular convolutional auto-encoder for
	channel coding,'' in \emph{Proc. IEEE 20th Int. Workshop Signal Process. Adv.
		Wireless Commun. (SPAWC)}, 2019, pp. 1--5.
	
	\bibitem{nachmani2016learning}
	E.~Nachmani, Y.~Be'ery, and D.~Burshtein, ``Learning to decode linear codes
	using deep learning,'' in \emph{Proc. 54th Annu. Allerton Conf. Commun.,
		Control, Comput. (Allerton)}.\hskip 1em plus 0.5em minus 0.4em\relax IEEE,
	2016, pp. 341--346.
	
	\bibitem{gruber2017deep}
	T.~Gruber, S.~Cammerer, J.~Hoydis, and S.~ten Brink, ``On deep learning-based
	channel decoding,'' in \emph{Proc. 51st Annu. Conf. Inf. Sci. Syst.
		(CISS)}.\hskip 1em plus 0.5em minus 0.4em\relax IEEE, 2017, pp. 1--6.
	
	\bibitem{nachmani2018deep}
	E.~Nachmani, E.~Marciano, L.~Lugosch, W.~J. Gross, D.~Burshtein, and
	Y.~Be’ery, ``Deep learning methods for improved decoding of linear codes,''
	\emph{IEEE J. Sel. Topics Signal Process.}, vol.~12, no.~1, pp. 119--131,
	2018.
	
	\bibitem{vasic2018learning}
	B.~Vasi{\'c}, X.~Xiao, and S.~Lin, ``Learning to decode {LDPC} codes with
	finite-alphabet message passing,'' in \emph{Proc. Inf. Theory Appl. Workshop
		(ITA)}.\hskip 1em plus 0.5em minus 0.4em\relax IEEE, 2018, pp. 1--9.
	
	\bibitem{teng2019low}
	C.-F. Teng, C.-H.~D. Wu, A.~K.-S. Ho, and A.-Y.~A. Wu, ``Low-complexity
	recurrent neural network-based polar decoder with weight quantization
	mechanism,'' in \emph{Proc. IEEE Int. Conf. Acoust., Speech, Signal Process.
		(ICASSP)}, 2019, pp. 1413--1417.
	
	\bibitem{buchberger2020prunin}
	A.~Buchberger, C.~H{\"a}ger, H.~D. Pfister, L.~Schmalen, and A.~G. Amat,
	``Pruning neural belief propagation decoders,'' in \emph{Proc. IEEE Int.
		Symp. Inf. Theory (ISIT)}, 2020, pp. 338--342.
	
	\bibitem{xu2017improved}
	W.~Xu, Z.~Wu, Y.-L. Ueng, X.~You, and C.~Zhang, ``Improved polar decoder based
	on deep learning,'' in \emph{Proc. IEEE Int. Workshop Signal Process. Syst.
		(SiPS)}, 2017, pp. 1--6.
	
	\bibitem{cammerer2017scaling}
	S.~Cammerer, T.~Gruber, J.~Hoydis, and S.~Ten~Brink, ``Scaling deep
	learning-based decoding of polar codes via partitioning,'' in \emph{Proc.
		IEEE Global Commun. Conf. (GLOBECOM)}, 2017, pp. 1--6.
	
	\bibitem{bennatan2018deep}
	A.~Bennatan, Y.~Choukroun, and P.~Kisilev, ``Deep learning for decoding of
	linear codes-a syndrome-based approach,'' in \emph{Proc. IEEE Int. Symp. Inf.
		Theory (ISIT)}, 2018, pp. 1595--1599.
	
	\bibitem{doan2018neural}
	N.~Doan, S.~A. Hashemi, and W.~J. Gross, ``Neural successive cancellation
	decoding of polar codes,'' in \emph{Proc. IEEE Int. Workshop Signal Process.
		Adv. Wireless Commun. (SPAWC)}, 2018, pp. 1--5.
	
	\bibitem{elias1954error}
	P.~Elias, ``Error-free coding,'' \emph{Research Laboratory of Electronics,
		Massachusetts Institute of Technology}, 1954.
	
	\bibitem{pyndiah1998near}
	R.~M. Pyndiah, ``Near-optimum decoding of product codes: Block turbo codes,''
	\emph{IEEE Trans. Commun.}, vol.~46, no.~8, pp. 1003--1010, 1998.
	
	\bibitem{mukhtar2016turbo}
	H.~Mukhtar, A.~Al-Dweik, and A.~Shami, ``Turbo product codes: Applications,
	challenges, and future directions,'' \emph{IEEE Commun. Surveys Tuts.},
	vol.~18, no.~4, pp. 3052--3069, 2016.
	
\end{thebibliography}
\end{document}